\documentclass[letterpaper]{jpconf}
\pdfoutput=1
\usepackage{graphicx}
\usepackage[square,sort&compress]{natbib}
\usepackage{amsmath}
\usepackage{amssymb}

\begin{document}
\title{Searching for \textit{hep} Neutrinos using the Sudbury Neutrino Observatory}

\author{Chris Howard, SNO Collaboration}

\address{University of Alberta\\ Edmonton AB\\
T6G 2G7}

\ead{choward@phys.ualberta.ca}

\begin{abstract}
The Sudbury Neutrino Observatory has recently finished its third and final phase, and has accumulated over 1082 days of neutrino data, spanning the energy range from approximately 5-20 MeV.  Almost all the observed neutrinos are due to the $^8$B reaction in the Sun.   The so-called \textit{hep} process ( $^3$He + p $\rightarrow$ $^4$He + e$^+$ + $\nu_e$) also occurs in the Sun, but has not yet been observed. The \textit{hep} neutrino energy endpoint extends above the $^8$B spectrum.  This paper describes the three phase analysis that will ultimately be the most sensitive to this reaction.

\end{abstract}

\section{SNO Overview}
\label{sno}

The Sudbury Neutrino Observatory is located near Sudbury, Ontario,
Canada, two kilometres underground in Vale INCO's Creighton mine. This
location uses the 6000 metre water equivalent (two km of norite rock) overhead as a shield against cosmic rays.  The detector itself is a
one-kilotonne ultra-pure heavy water Cerenkov light detector\cite{snonim}. The
heavy water (D$_2$O) is contained in a 12 metre diameter spherical acrylic
vessel (AV), surrounded by a 17.8 metre geodesic
photomultiplier support structure (PSUP). This frame houses nearly
10,000 photomultiplier tubes. To shield the heavy water from
external backgrounds and help support the AV the remainder of the cavity is filled with light water. 

Neutrinos can exist in three probability states known as \textit{flavours}\cite{bahcall}. SNO has the unique ability to detect all three neutrino flavours
via the reactions:

\begin{equation}
\nu_e + d \rightarrow p + p + e^-
 \label{Eq:CC}
\end{equation}
\begin{equation}
\nu_x + d \rightarrow p + n + \nu_x
 \label{Eq:NC}
\end{equation}
\begin{equation}
\nu_x + e^- \rightarrow \nu_x + e^-
 \label{Eq:ES}
\end{equation}

These equations describe the charged current (CC), neutral current
(NC) and elastic scattering (ES)
reactions, respectively. The charged current reaction is only
sensitive to the $\nu_e$ flavour, but the neutral current reaction
is sensitive to any of the three flavours, $\nu_x$, where $x$ can
be $e$, $\mu$, or $\tau$. All prior solar neutrino experiments showed a deficit of
$\nu_e$ neutrinos to the predicted rate by the standard solar model (SSM)\cite{homestake}\cite{gallex}\cite{sage}\cite{kamioka}. SNO's sensitivity to both NC and CC allowed SNO
to distinguish an oscillation scenario ($\frac{\nu_e}{\nu_x} < 1$) from a solar model scenario ($\frac{\nu_e}{\nu_x} =
1$). SNO solved the {\sl solar neutrino problem} by showing that
the rate of NC reaction events agreed with the number predicted by the standard solar
model\cite{sno1}, while the rates of CC reaction
events and ES events were consistent with
oscillations in which two-thirds of the neutrinos at the Earth are
not $\nu_\e$ neutrinos.

The SNO experiment was done in three phases. Each phase employed a
different technique to detect neutrons. The first phase looked for
neutron capture on the deuterium\cite{sno1}. The second phase involved the
addition of NaCl salt to the active volume\cite{sno2}. Chlorine has a larger cross section for neutron capture and thus improved the efficiency of the NC reactions and this increased the statistical precision. The third and final phase of SNO removed the NaCl and added $^3$He
proportional counters in the active volume\cite{sno3}. The counters were referred to as neutral current detectors (NCDs).  The advantages were to increase
the efficiency for the detection of the NC reaction,
as compared to pure D$_2$O, and to allow event-by-event particle
identification.

\begin{figure}[!htp]
\centering
\includegraphics[scale=0.7]{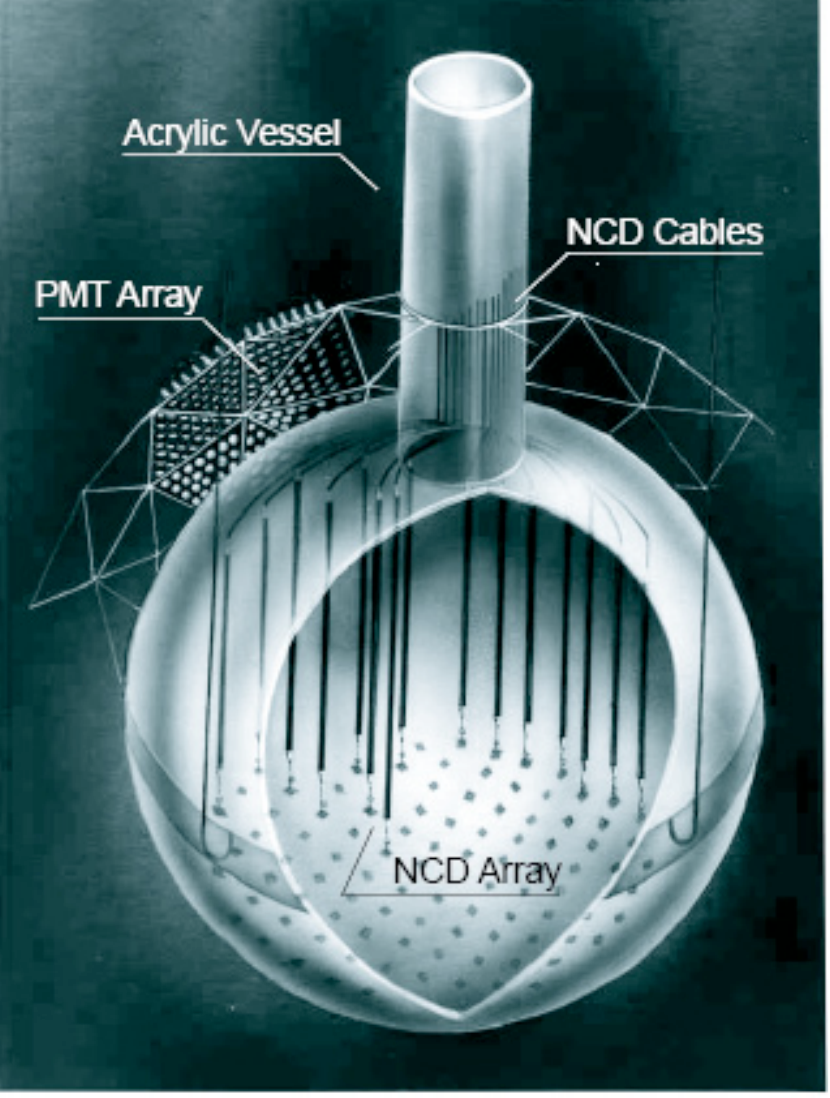}
\caption[NCD Array]{An artist's drawing of SNO and the NCD array.}
\label{NCDarray}
\end{figure}

\section{The \textit{hep} Neutrinos}

\begin{figure}[htb]
\begin{minipage}{18pc}
\includegraphics[width=14pc, angle=-90]{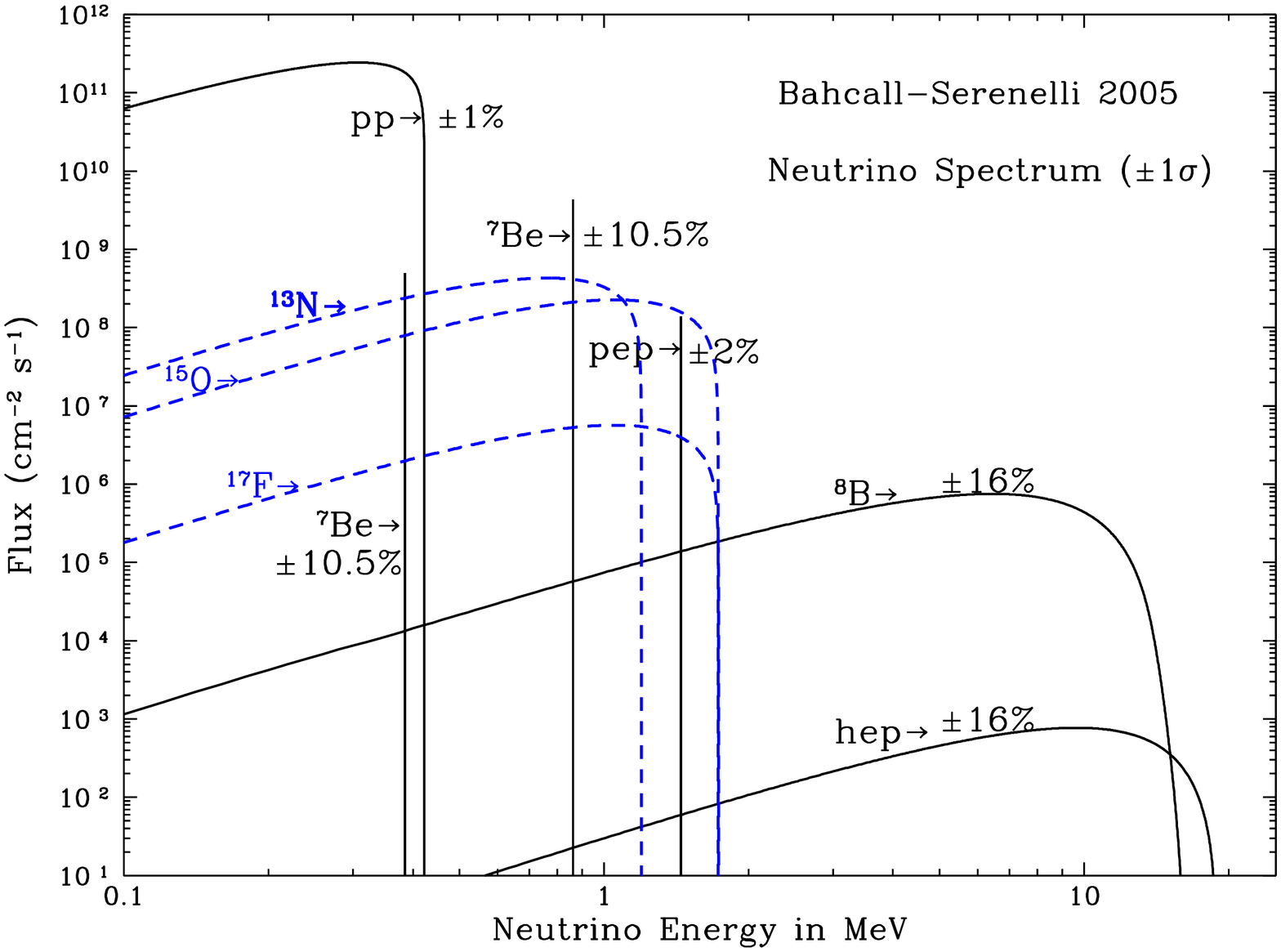}
\caption{\label{solarspec}Solar neutrino energy spectra.\cite{bahcall-2005}}
\end{minipage}\hspace{2pc}%
\begin{minipage}{16pc}
\vspace{2pc}
\includegraphics[width=18pc]{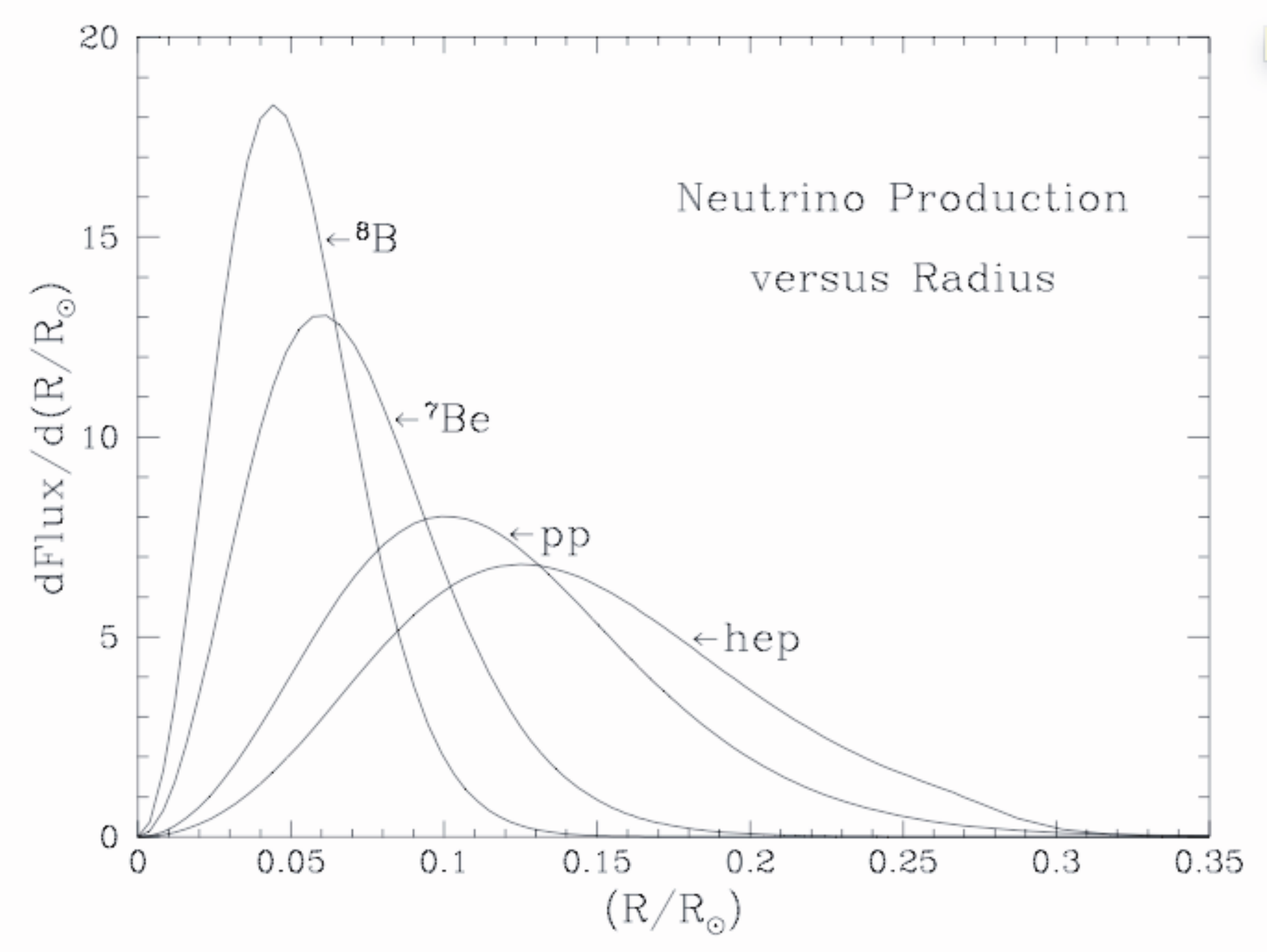}
\caption{\label{solarradius}Radial profile of neutrino production in the Sun.\cite{bahcall}}
\end{minipage}
\end{figure}

The Sun produces neutrinos through many reactions, and their respective energy spectra are shown in Figure \ref{solarspec}. The \textit{hep} reaction extends to the highest energy, but is rare compared to the $^8$B signal which overwhelms the \textit{hep} signal until near 15 MeV. The \textit{hep} neutrinos are produced at a larger radius than other solar neutrinos, as shown in Figure \ref{solarradius}, and can therefore test different parameters of the solar model.

\section{The \textit{hep} Analysis}

The SNO experiment was designed to search for solar neutrinos mainly from the $^8$B reaction. This has been done in each of the three phases as discussed is Section \ref{sno}. The \textit{hep} reaction has never been detected and falls in SNOs accessible energy region, so a search should be conducted. SNO has previously searched using only the first phase of data, and sets the best limit on the \textit{hep} neutrino flux \cite{heppaper}. The first SNO \textit{hep} search was conducted using a \textit{box analysis} which was a simple counting experiment in a predefined energy region with no spectral information used. Our analysis will be a spectral analysis.

The \textit{hep} neutrino analysis must be able to use the high energy tail from the \textit{hep} energy spectrum to distinguish \textit{hep} neutrinos from $^8$B neutrinos. It must also be able to permit for a signal extraction when one or more of the signals have very few events. We will use Markov Chain Monte Carlo (MCMC) with a Metropolis-Hastings algorithm \cite{hastings} to extract the number of \textit{hep} neutrinos. This will be done using all three phases of the SNO data set, totaling about 1082 days of livetime.

\begin{figure}[htbp]
\begin{center}
\includegraphics[scale=0.57, angle=90]{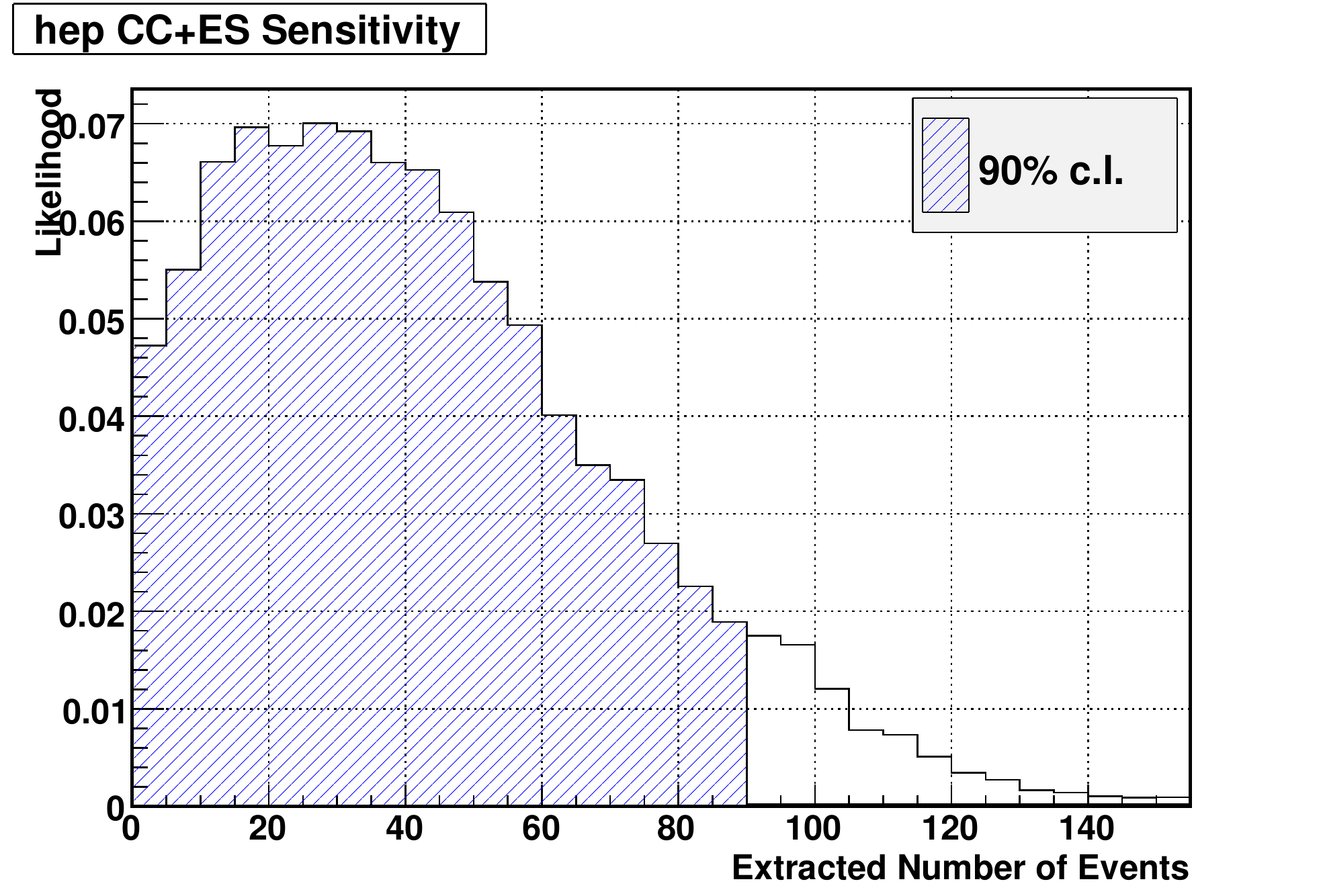}
\caption[]{Sensitivity to 19 \textit{hep} events using Monte Carlo as input data for a sample signal extraction.}
\label{fig:sens}
\end{center}
\end{figure}

The MCMC method will be applied in a Bayesian analysis, so a prior, $\pi(\theta)$, needs to be applied. We chose the prior shown in Equation \ref{eq:prior}

\begin{equation}
\pi(\theta) = 
\begin{cases} 
  0  & \mbox{if }\theta<0 \\
  1  & \mbox{if }\theta\ge0 
\end{cases}
\label{eq:prior}
\end{equation}

\noindent where $\theta$ is any signal in the fit. This prior returns a posterior distribution that represents the likelihood. In other words, the MCMC will \textit{map out} the likelihood space. This has the advantage of producing a parameter that cannot be negative and allows us to easily set a 90\% confidence limit. Equation \ref{eq:cl} shows how we will define the confidence limit\cite{cowan}:

\begin{equation}
1-\beta = \int_{-\infty}^{\theta_{up}}p\left(\theta|x\right)d\theta
\label{eq:cl}
\end{equation}

\noindent where $1-\beta$ is $0.90$ for a 90\% confidence limit, $p\left(\theta|x\right)$ is the posterior distribution as determined by the MCMC method and $\theta_{up}$ is the number of events at the desired confidence limit. Figure \ref{fig:sens} shows this integration and was created by running the signal extraction over a data set made up of Monte Carlo events.

\section{Conclusion}

Our analysis has an improvement of more than a factor of three in statistics, meaning more than a factor of $\frac{1}{\sqrt{3}}$ in our expected sensitivity. Our analysis also takes into account the shape of the energy spectrum which was lost in the box technique. This will be the best limit on the \textit{hep} flux for SNO.
 
\bibliography{choward_LLWI09}
\bibliographystyle{iopart-num}

\end{document}